\documentstyle[pre,preprint,aps]{revtex}
\tightenlines

\newcommand{\simg}{\stackrel{>}{\sim}}

\begin{document}
\draft

\title{
An associative memory of Hodgkin-Huxley neuron networks
with Willshaw-type synaptic couplings
\footnote{E-print: cond-mat/0007198} 
}
\author{
Hideo Hasegawa\footnote{E-mail: hasegawa@u-gakugei.ac.jp}
}
\address{
Department of Physics, Tokyo Gakugei University,
Koganei, Tokyo 184, Japan
}
\date{\today}
\maketitle
\begin{abstract}

\end{abstract}
An associative memory has been discussed of neural
networks consisting of spiking $N \:(=100)$ Hodgkin-Huxley (HH) neurons
with time-delayed couplings, which
memorize $P$ patterns in their synaptic weights.
In addition to excitatory synapses whose strengths are modified
after the Willshaw-type learning rule with the 0/1 code
for quiescent/active states, the network includes uniform
inhibitory synapses which are introduced to reduce cross-talk noises. 
Our simulations of the HH neuron network for
the noise-free state have shown to yield a fairly good performance with
the storage capacity of $\alpha_c = P_{\rm max}/N \sim 0.4 - 2.4$
for the low neuron activity of 
$f \sim 0.04-0.10$. 
This storage capacity of our temporal-code network 
is comparable to that of 
the rate-code model with the Willshaw-type synapses.
Our HH neuron network is realized not 
to be vulnerable to
the distribution of time delays in couplings. 
The variability of interspace interval (ISI) of output
spike trains in the process of retrieving stored patterns is also discussed.

\vspace{1.0cm}
\pacs{PACS No. 84.35.+i  87.10.+e  87.18.Sn}
%

\begin{center}
{\bf I. INTRODUCTION}
\end{center}

The Hopfield-type neural networks have been intensively 
investigated as a model for the learning and memory of brain 
\cite{Hopfield82}.
In this type of models the information is stored as 
a content-addressable memory in 
which synaptic strengths are modified after 
the Hebbian rule.
Its retrieval is made when the network with the symmetric couplings
works as the point-attractor with the fixed points.
These models have been analyzed by the method of statistical
mechanics of spin glasses \cite{Amit85}.
Such analysis provides us with most important information of
storage capacity, role of noise and recall performance.
The Hopfield-type models, however, are
an extreme abstraction of real neural networks and
they have the following issues from the biological viewpoint.

(i) The dynamical variable in the Hopfield-type model
is the firing rate of neurons ({\it rate coding}). However,
the importance of a precise timing of firing ({\it temporal coding})
is currently realized in many experiments:
sonar processing of bats \cite{Suga83}, 
sound localization of owls \cite{Konishi92},
electrosensation in electric fish \cite{Carr86}, 
visual processing
of cats \cite{Eckhorn88} \cite{Gray89}, 
monkeys \cite{Rolls94} and human \cite{Thorpe96}.

(ii) The synaptic weight $W_{jk}$ for a given synapse
between the pre-synaptic neuron $k$ and the
post-synaptic neuron $j$ in the Hopfield-type models
is usually assumed to be given by\cite{Hopfield82}
\begin{equation}
W_{jk} = \sum_{\mu=1}^P 
(\xi_j^{(\mu)}-a) \: (\xi_k^{(\mu)}-b),
\end{equation}
where $P$ stands for the number of patterns to be stored,
$\xi_j^{(\mu)}$= -1/1 (or 0/1) for quiescent/active states 
and $a=b=f  \equiv <\xi_j^{(\mu)}>$, the average activity.
Equation (1) shows that a given synapse may be positive 
(excitatory) or negative (inhibitory) depending
on the covariance of the pre- and post-synaptic activities.
This is not biologically realized because the synapse is
either excitatory or inhibitory with little exception.
Furthermore, Eq.(1) shows that 
the potentiation of the synaptic strength of a given synapse 
is possible even in the absence of pre- and post-synaptic activities
($\xi_j^{(\mu)}=\xi_k^{(\mu)}=-1$ or 0).
The potentiation without the brain activity again contradicts the
biological evidence.

In recent years
some theoretical studies have been reported of the learning and 
memory in terms of the temporal coding
with the use of spiking neurons \cite{Gerstner92}-\cite{Kanamaru00}
or oscillating neurons \cite{Aonishi99}-\cite{Yoshioka00}.
It has been shown that spiking neurons
such as the spike-response and integrate-and-fire neurons 
with a correlation-based Hebbian learning rule can
operate very fast in the order of ten milliseconds
\cite{Gerstner92} \cite{Treves97}.
Mueller and Herz \cite{Mueller99} showed that the networks consisting of 
spiking integrate-and-fire (IF) neurons possess the storage
capacity similar to those of rate-code neurons.
Their model, however, adopts the leaning rule given 
by Eq.(1) for storing patterns in synapse weights, 
and then it has the second issue raised above.
Based on spiking FitzHugh-Nagumo (FN) neurons
with time-delayed couplings, Yoshioka and Shiino \cite
{Yoshioka98}, and Kanamuru and Okabe \cite{Kanamaru00} 
discussed the associative memory with the use of the
leaning rule given by Eq.(1) but with $a = 0$ and $b=f$
for the 0/1 code.
Their approach has the second issue because their $W_{jk}$
may be positive for negative depending on whether 
$\xi_k^{(\mu)}=1$ or 0 for $\xi_j^{(\mu)}=1$.

Among spiking neuron models having been proposed so far, 
the Hodgkin-Huxley (HH) model
is expected to be the most realistic in the biological sense.
Since the HH model was proposed in 1952 \cite{Hodgkin52}, 
its property has been intensively investigated.
Its responses to applied dc \cite{Nemoto76} 
and sinusoidal currents \cite{Aihara84}, and
spiking impulses \cite{Hasegawa00} have been studied.
The HH model has been generalized with modifications
in ion conductances and widely adopted for a study
of biological systems such as hippocampus \cite{Hoppocampus} 
and thalamus \cite{Thalamus}. 

Quite recently, Lytton \cite{Lytton98} has proposed
the feedfoward hetero-associative memory
network of spiking 40 HH neurons, by employing
the notion of the Anderson-Kohonen network.
The weight of the excitatory synapses is assumed to
be given by the conventional Hebbian rule:
\begin{equation}
W_{jk} = \sum_{\mu=1}^P 
\xi_j^{(\mu)} \: \xi_k^{(\mu)},
\end{equation}
with $\xi_j^{\mu}=0/1$ for quiescent/active states.
In addition to the excitatory synapses,
Lytton has included inhibitory synapses whose weights are determined
by a feedfoward rule depending on input patterns, 
in order to suppress the cross-talk noises arising from the
non-orthogonal inputs.
Equation (2) shows that the synapse is strengthen only when
both pre- and post-synapses are active, in agreement with
biological experiments.
It is shown, however, that the retrieval of the HH network
is only possible in the narrow parameter ranges of
conductances of excitatory and inhibitory synapses \cite{Lytton98}.
 
Since a naive transplant of the Hebbian rule given by Eq.(2)
to HH neuron networks yields the poor performance in retrieving 
memorized patterns, as was shown by Lytton \cite{Lytton98},
we adopt, in this paper, the Willshaw rule given by \cite{Willshaw69}
\begin{equation}
W_{jk} = \: \Theta
(\sum_{\mu=1}^{P} \: \xi_{j}^{(\mu)} \: \xi_{k}^{(\mu)}),
\end{equation}
where $\Theta(x)=1$ for $x>0$ and 0 otherwise,
and $\xi_j^{(\mu)}=0/1$ for quiescent/active states.
The synaptic weight given by Eq.(3) is either one or zero,
in contrast with that given by Eqs. (1) and (2).
Such a dramatic reduction is not biologically justifiable 
except under some conditions when the postsynaptic 
membrane of a dendritic spine is active \cite{Kohring90}.
It has been, however, reported that the Willshaw rule given by Eq.(3)
improves the storage efficiency of
Hopfield-type neural networks 
\cite{Willshaw69}-\cite{Brunel94}.
Equation (3) shows that the synaptic weight is
potentiated only for the simultaneous firings of
the pre- and post-synaptic neurons, which is in accord with
biological results.

Theoretical studies on memory have extensively 
pursued the analytical approach, 
requiring simplified tractable models, which do not necessarily
reflect the aspect of the biological reality.
As an alternative approach, we may take a simulation method for 
a better understanding of functioning of real, biological systems
to which an analytical method cannot applied.

It is the purpose of the present paper
to construct biologically plausible model for an associative memory.
We propose HH neuron networks with excitatory and inhibitory synapses, 
weights of the former are modified by the Willshaw-type learning rule.
Simulations of the memory function of the network are performed
for sparse coding with the low firing activity ($f \ll 1$)
because neurophysical evidences indicate that only a small fraction
of neurons are active at a given time \cite{Note1}.
Our simulations demonstrate
that our HH neuron networks
works fairly well in memory function within the temporal coding.

Our paper is organized as follows:
In the next Sec. II, we describe a neural system
consisting of $N$ HH neurons which are fully connected
with time-delayed couplings.
The results of simulations are presented in Sec. III.
The retrieval of the stored patterns
by  perfect and imperfect input patterns are studied
in Sec. 3.1 and Sec. 3.2, respectively, and the effect of the
distribution of the time delay in couplings 
is discussed in Sec. 3.3.  
The final Sec. IV is devoted to discussions concerning the relevant
rate-code model and the variability of interspike intervals (ISI) of 
output spike trains.

\begin{center}
{\bf II. ADOPTED NETWORK MODEL}
\end{center}

We adopt a network of $N$ HH neurons
which stores and retrieves sparsely coded $P$ patterns.
HH neurons described by
identical parameters,
are fully coupled by synapses with time-delayed couplings.
The input information is assumed to be stored in the
synaptic plasticity as will be discussed shortly [Eq.(11)].

Dynamics of the membrane potential $V_{j}$ of
the HH neuron {\it j} $(=1,..,N)$ 
is described by the non-linear delay-differential
equations given by 

\begin{equation}
C \: d V_{j}(t)/d t = -I_{j}^{\rm ion}(V_{j}, m_{j}, h_{j}, n_{j})  
+ I_{j}^{\rm ext} + I_{j}^{\rm int},
\end{equation}
where $C = 1 \; \mu {\rm F/cm}^2$ is the capacity of the membrane.
The first term of Eq.(4) expresses the ion current given by
\begin{equation}
I_{j}^{\rm ion}(V_{j}, m_{j}, h_{j}, n_{j}) 
= g_{\rm Na} m_{j}^3 h_{j} (V_{j} - V_{\rm Na})
+ g_{\rm K} n_{j}^4 (V_{j} - V_{\rm K}) 
+ g_{\rm L} (V_{j} - V_{\rm L}).
\end{equation}
Here the maximum values of conductivities 
of Na and K channels and leakage are
$g_{\rm Na} = 120 \; {\rm mS/cm}^2$, 
$g_{\rm K} = 36 \; {\rm mS/cm}^2$ and
$g_{\rm L} = 0.3 \; {\rm mS/cm}^2$, respectively,
and the respective reversal potentials are   
$V_{\rm Na} = 50$ mV, $V_{\rm K} = -77$ mV and 
$V_{\rm L} = -54.5 $ mV.
The dynamics of gating variables of Na and
K channels, $m_{j}, h_{j}$ and $n_{j}$,
are described by linear-differential equations,
whose explicit expressions have been given 
(for example see \cite{Hasegawa00}).

The second term in Eq.(4) denotes the external input current given by
\begin{equation}
I_{j}^{\rm ext} = g_{syn} \: (V_a-V_c) \: \sum_{n} \alpha(t-t_{in}),
\end{equation}
which is induced by
the  pre-synaptic spike-train input applied 
to the neuron $i$, given by
\begin{equation}
U_{\rm i}(t) = V_{\rm a} \: \sum_n  \: \delta (t - t_{{\rm i}n}).
\end{equation}
In Eqs.(6) and (7), $t_{{\rm i}n}$ is the $n$-th firing time of 
the spike-train inputs, $g_{syn}$ and $V_c$ denote the conductance 
and the reversal potential, respectively, of the synapse,  
$\tau_{\rm s}$ is the time constant relevant 
to the synapse conduction,
and $\alpha(t)$ is 
the alpha function  given by
\begin{equation}
\alpha(t) = (t/\tau_{\rm s}) \; e^{-t/\tau_{\rm s}} \:  \Theta(t).
\end{equation}
where $\Theta (t)$ is the Heaviside function.

When the membrane potential of the {\it j}-th neuron $V_{j}(t)$ oscillates,
it yields the spike-train output, which may be expressed by
\begin{equation}
U_{{\rm o}j}(t) = V_{\rm a} \: \sum_m \: \delta (t - t_{{\rm o}jm}),
\end{equation}
in a similar form to Eq.(7), 
$t_{{\rm o}jm}$ being the $m$-th firing time 
when $V_{j}(t)$ crosses $V_{z}$ = 0.0 mV from below.

The third term in Eq.(4) 
expresses the interaction between neurons:
\begin{equation}
I_{j}^{\rm int}
=   G( \sum_{k (\neq j)} \sum_{m} 
\: (g_{jk}^{exc}- g_{jk}^{inh}) (V_a-V_c)
\: \alpha(t-\tau_{jk}-t_{{\rm o}km})).
\end{equation}
with
\begin{equation}
g_{jk}^{exc} = g_{exc} \: \Theta
(\sum_{\mu=1}^{P} \: \xi_{j}^{(\mu)} \: \xi_{k}^{(\mu)}),
\end{equation}
\begin{equation}
G(x)= x \: \Theta(x),
\end{equation}
In Eqs.(10)-(12) the self-interaction ($j=k$) is forbidden,
$g_{jk}^{exc}$ denotes the conductance 
of the excitatory synapse, 
$g_{jk}^{inh}$ that of the inhibitory synapse,  and
the time delay $\tau_{jk}\:(=\tau_{kj}=\tau_d)$ is 
the sum of conduction times for currents from the neurons $k$ to $j$
through the axon and dendrite. 
The excitatory synaptic strength $g_{jk}^{exc}$ in Eq.(11)
takes either the value of 0 or $g_{exc}$
after Willshaw {\it et al.} \cite{Willshaw69}\cite{Golomb90}.
The synaptic plasticity occurs only when both 
the pre- and post-synaptic neurons fire simultaneously,
in accord with biological findings.
The uniform inhibition $g_{jk}^{inh}=g_{inh}$ is included to
reduce the cross-talk noise. 

The function $G(x)$ in Eq.(12) represents the dendrite processing
of postsynaptic currents,
and its uni-directional form is adopted such that
neurons do not fire for resultant inhibitory inputs \cite{Note2}.
It is well known that an HH (and FN) neuron can fire 
even for an inhibitory input with the
inhibitory rebound process \cite{Perkel74}, 
whereas an IF neuron cannot.
Figure 1(a) and 1(b) show responses of a single HH neuron to
excitatory and inhibitory inputs, respectively,
which are calculated by setting 
$I_j^{int}=0$ in Eq.(4).
With an application of an excitatory input,
a HH neuron is depolarized to fire at 2.8 msec
after the trigger, and it recovers to the rest state
with the refractory period of about 10 msec.
When an inhibitory input is applied, on the other hand,
a HH neuron is once hyper-polarized and then depolarized  
to fire after 14.6 msec of the input injection.
Because the firing of neurons for inhibitory inputs is not
appropriate for the memory retrieval, we introduce 
the uni-directional function $G(x)$ given by Eq.(12), 
which suppresses the negative (inhibitory) input, 
related discussions being given in Sec.IV.

Patterns to be stored
are expressed by a vector of 
{\boldmath $\xi^{(\mu)}$}
$=\{\xi_{j}^{(\mu)} \: \mid \: j=1-N \}\: $ 
where $\mu=1-P$ and $\xi_{j}^{(\mu)}= 0 \: (1)$ 
for the quiescent (firing) state of the neuron $j$.
We randomly create the patterns to be stored such as
to satisfy the condition for a given number of unit-codes,
$M\:(=fN)$:
\begin{equation}
M = M^{(\mu)}= \sum_j \xi_j^{(\mu)}.
\end{equation}
Without a loss of the generality, we take 
for the $\mu=1$ pattern that
$\xi_{j}^{(1)} =1$ for $j \leq M$ and 
$\xi_{j}^{(1)} =0$ for $j > M$.
Simulations have been repeated
for ten samples, each of which includes $P$ patterns.
The standard deviations in the calculated results 
are shown by error bars in figures
of the following sections.

The retrieval of the first pattern of
{\boldmath $\xi^{(1)}$} is made by injecting
a proper input pattern of {\boldmath $\zeta$}
to our network at $t=0$.
The input pattern may be 
{\boldmath $\zeta=\xi^{(1)}$} (perfect input)
or {\boldmath $\zeta \neq \xi^{(1)}$} (imperfect input).
Differential equations given by Eqs.(4)-(12) are solved 
by the forth-order Runge-Kutta method
by the integration time step of 0.01 msec
with double precision.
We assume that all HH neurons are in the same
initial conditions given by
\begin{equation}
V_{j}(t)= -65 \:\: \mbox{\rm mV}, m_{j}(t)=0.0526,  
h_{j}(t)=0.600, 
n_{j}(t)=0.313, \:\: \mbox{\rm for} \:\: j=1, 2
\:\:  \mbox{at} \: t=0,
\end{equation}
which are the rest-state solution of a single
HH neuron ($I_j^{int}=0$).
The initial function for $V_{j}(t)$ of all HH neurons
is given by
\begin{equation}
V_{j}(t)= -65 \:\: \mbox{\rm mV} 
\:\: \mbox{\rm for} \:\: j=1,2
\:\:  \mbox{at} \: t \in [-\tau_d, 0).
\end{equation}

The adopted model parameters are summarized as follows: $N=100$,
$V_a=30$ mV, $V_c=-50$ mV, $\tau_d=10$ msec,
$\tau_{\rm s}=2$ msec \cite{Hasegawa00},
$g_{syn}=\:g_{exc}=0.3 \; {\rm m S/cm}^2$ 
and $g_{inh}=0.24 \; {\rm m S/cm}^2$.
The time delays are taken to be a single constant value of
$\tau_{jk}=\tau_{kj}=\tau_{d}$ in Sec. 3.1 and 3.2, 
while in Sec. 3.3
they are allowed to uniformly distribute in the limited range.
We choose $\tau_d=10$ msec such that it not only satisfies the condition of
$\tau_d > 7$ msec, which arises from the absolute refractory
period of a HH neuron [Fig. 1], but also it is witnin 
the biologically conceivable values \cite{Yoshioka98}. 
A choice of $g_{exc}$ and $g_{inh}$ will 
be discussed shortly.
Although the size of $N$ (and $P$ and $M$) is not sufficient large 
because of a limitation of our computer facility, the calculated 
results are expected to be useful and meaningful.

\vspace{0.5cm}
\begin{center}
{\bf III. RESULTS OF SIMULATIONS}
\end{center}

\begin{center}
{\bf 3.1 Retrieval by perfect inputs}
\end{center}


Firstly we consider, in this subsection, the case in which 
{\boldmath $\zeta=\xi^{(1)}$} is injected for retrieval 
of {\boldmath $\xi^{(1)}$} (the perfect input):
the retrieval by imperfect-input injection 
({\boldmath $\zeta \neq \xi^{(1)}$})
will be
discussed in the following subsection 3.2. 

We define
the initial overlap, $m_i$, between the input pattern {\boldmath $\zeta$}
and the pattern to be retrieved {\boldmath $\xi^{(1)}$}:
\begin{equation}
m_i = (1/N) \sum_j\: [2 \xi_j^{(1)}-1]\: [2 \zeta_j-1].
\end{equation}
It is easy to see that $m_i=1.0$ for $\zeta_j=\xi_j^{(1)}$.
Similarly the time-dependent overlap, $m(t)$,
between the state of the system
and the pattern {\boldmath $\xi^{(1)}$}, is defined by
\begin{equation}
m(t) = (1/N) \sum_j\: [2 \xi_j^{(1)}-1]\: [2 \eta_j(t)-1],
\end{equation}
where $\eta_j(t)$ is given by
\begin{eqnarray}
\eta_j(t)&=&1, \;\;\; \mbox{ if $U_{oj}=1$ at 
$t \in [t_{ojm}-\delta t,\:t_{ojm}+\delta t]$ }, \nonumber \\
&=&0, \;\;\;\;\;\;\;\;\;\;\;\;\;\;\;\;\;\mbox{otherwise},
\end{eqnarray}
$t_{{\rm o}jm}\; (m=1,2,..)$ being the firing time of the neuron $j$
and $\delta t \;(= 2.45 \:\tau_{syn} \sim 5$ msec) denoting
the margin with a magnitude of
a half-width of the alpha function given by Eq.(8).
We hereafter regard firings with
the overlap of $m(t)=1.0$ at $t \simg 50 \: \tau_d$ 
as a {\it successful retrieval}. 

An example of an retrieval of
of {\boldmath $\xi^{(1)}$} for a given set of patterns, 
is shown in Fig. 2. We show
the time course of a trigger input $U_{ij}$,
the output $U_{oj}$,
the pre-synaptic current $I_j\;(=I_j^{ext}+I_j^{int})$ and
the membrane potential $V_j$
of the neuron of $j=1$,
for an injection of the input pattern {\boldmath $\zeta$}
at $t=0$ with $\alpha=0.30$, $f=0.1$, 
$g_{sys}=g_{exc}=0.30$ and $g_{inh}=0.24 \; {\rm m S/cm}^2$.
The pre-synaptic current $I_1^{ext}$ starting from $t=0$
is yielded by the trigger input $U_{{\rm i}j}$,  
while other pre-synaptic currents, for example,
starting from $t=12.6$ msec, arise from the interaction term $I_j^{int}$
through the couplings.
The firing of the neuron $j=1$ is given by $U_{o1}(t)=1$, and
it is compactly depicted in Fig. 3(a), where
dots express firings of neurons numbered by 
the index $j= 1-100$ in the vertical scale
as a function of time [$U_{oj}(t)=1$ in Eq.(9)].
Figure 3(a) shows a successful retrieval for $\alpha=P/N=0.30$ with
$f=M/N=0.10$, $g_{exc}=0.30\; {\rm m S/cm}^2$ and $\lambda=0.8$.
The time-dependent overlap $m(t)$ given by Eq.(17)
is always unity: $m_i=m(t)=1.0$ at $t \geq 0$,
which is calculated at 
$t=t_{{\rm o}1m}\; (m=1, 2,..)$,
firing times of the neuron 1, and
which is plotted at the upper part of Fig. 3(a).
The period of firings $T_{o}$ is determined by
$T_{o} = \tau_d + \tau_r \; \sim 12.5$ msec 
where $\tau_d$ denotes the delay time of couplings and
$\tau_r \; \sim 2.5$ msec is the response time of a HH neuron.
Figure 3(b), on the contrary, shows a failed case for 
a larger $\alpha$ value of $\alpha=0.50$, yielding extra firings of
neurons of $j > 10$ and non-firings of $j \leq 10$.
We note that $m_i=m(t)=1.0$ at $t=0$ but $m(t)$ quickly deviates 
from its initial value at $t > 0$.
Figures 3(a) and 3(b) show that a single trigger impulse 
is sufficient to recall the stored pattern. 

In order to show the importance of the uniform, inhibitory synapse
introduced by $g_{inh}$ in Eq.(10), 
we plot, in Fig. 4,
the $\lambda \:(=g_{inh}/g_{exc})$ dependence of the critical 
storage capacity $\alpha_c = P_{\rm max}/N$, above which the retrieval
of the input pattern cannot made.
We note that $\alpha_c$ decreases as the value of
$\lambda$ decreases.
Although the $\lambda$ dependence of $\alpha_c$ with 
$g_{exc}=0.3 \; {\rm m S/cm}^2$ 
is slightly different from that with 
$g_{exc}=0.5\; {\rm m S/cm}^2$ for $f=0.06$,
the result with $g_{exc}=0.3 \; {\rm m S/cm}^2$ is identical with 
that of $g_{exc}=0.5 \; {\rm m S/cm}^2$ for $f=0.10$.
We have decided to adopt $g_{exc}=0.3 \; {\rm m S/cm}^2$ and $\lambda=0.8$
($g_{inh}=0.24 \; {\rm m S/cm}^2$)
for our simulations.

The calculated storage capacity 
$\alpha_c$ as a function of the mean activity $f$ is plotted 
in Fig. 5, where bars express the standard
deviations due to the sample dependence.
It clearly shows that $\alpha_c$ increases as $f$ decreases.
Theoretical analysis based on the Hopfield model has
shown that the storage capacity
diverges as $\alpha_c \propto (-1/f \;{\rm log } f)$ in the limit
of vanishing $f$ \cite{Gardner88}-\cite{Amari89}.
Recent calculations of oscillating associative memory
made by Aoyagi and Nomura \cite{Aoyagi99} also support
the $(-1/f \;{\rm log } f)$ dependence.
Although the $f$ values of our simulations for spiking 
HH neuron networks are not sufficiently small,
our result seems to support it:
the dotted curve expresses $10 /(M \:{\rm log} M)$, which is
proportional to $(-1/f \; {\rm log} f)$ in the limit of 
$f \rightarrow 0$.

\begin{center}
{\bf  3.2 Retrieval by imperfect inputs}
\end{center}

Next we discuss the retrieval of {\boldmath $\xi^{(1)}$} when an imperfect
input, {\boldmath $\zeta \neq \xi^{(1)}$},
is injected. We assume that an input pattern {\boldmath $\zeta$}
is similar to {\boldmath $\xi^{(1)}$} but some of its codes are
modified, keeping unit-code number unchanged: $M = \sum_j \zeta_j$.
We adopt {\boldmath $\zeta$} whose codes are same as those of {\boldmath $\xi^{(1)}$}
except 1-codes in $j \in [M-\Delta M/2, \: M)]$ are changed to 0-codes, and
0-codes in $j \in [M+1, M+\Delta M/2]$ to 1-codes , where 
even $\Delta M \:(\geq 2)$ is the number of changed codes. 
The initial overlap of {\boldmath $\zeta$} thus created is 
$m_i=1.0-2 \: \Delta M/N$. 

Figure 6(a) shows an example of
the successful retrieval
when an input with the initial overlap of
$m_i=0.84$ is injected for $\alpha=0.10$ and $f=0.20$.  
The time dependence of $m(t)$ plotted in Fig. 7,
shows that
$m(t)= m_i=0.84$ just after an input injection
and $m(t)$ gradually approaches unity.
The time dependence of $m(t)$
for the case of $m_i=0.92$ also successfully approaches unity.
On the contrary, when an input pattern has 
a smaller initial overlap of $m_i=0.68$, the retrieval
is failed; firings of neurons spread all over the system as shown
in Fig. 6(b) and the overlap $m(t)$ decreases from the
initial value of $m_i=0.68$ as shown in Fig. 7. 
The failed case is plotted also for $m_i=0.52$, whose
transient value of $m(t)$ is not necessary smaller
than that for $m_i=0.68$.

The dashed curves in Fig. 8 show the storage capacity 
$\alpha_c$ against the initial overlap $m_i$ for $f=0.20$.
It shows the critical, initial overlap is about $m_i=0.73$
for $\alpha=0.1$ and $f=0.20$ (Fig. 7).
We note in Fig. 8 that for $f=0.20$, the retrieval is
effectively possible for $m_i \simg 0.6$.
On the contrary, for $f=0.10$, the retrieval is possible 
only in the restricted region of $0.92 < m_i \leq 1.0$. 
Figures 7 and 8 reminds us the similar calculation made by
Amari and Maginu for the Hopfield model \cite{Amari88}.

\begin{center}
{\bf 3.3 Effect of distributed time delays}
\end{center}

So far we have assumed that the time delays 
are the same for all couplings connecting neurons:
$\tau_{jk}=\tau_{kj}=\tau_d$.
It is expected to be not the case in real biological systems.
In order to study the effect of their distributions,
we assume that time delays uniformly (randomly) distribute as
\begin{equation}
\tau_{jk}=\tau_{kj} \in [\tau_d-\Delta \tau/2,\; \tau_d+\Delta \tau/2],
\end{equation}
where $\Delta \tau$ is a width of the time-delay distribution.

When the width of the time-delays distribution becomes large, 
it is difficult for post-synaptic currents to sustain 
firings of neurons.
The solid curve in Fig. 9, expressing the critical storage $\alpha_c$
for $f=0.06$ as a function of $\Delta \tau$ for the perfect-input
retrieval, shows that
the critical storage decreases when the 
width of the distribution becomes wide, and that $\alpha_c$ vanishes for
$\Delta \tau \geq 6$ msec.  
In the case of $f=0.10$, whose
result is plotted by dashed curve, $\alpha_c$ vanishes for
$\Delta \tau \geq 9$ msec.  
Although the time-dependent overlap of $f=0.10$ deviates from 1.0
for $\Delta \tau > 9$ msec,
the oscillating firings may continue with spread
firing times.
Fig. 10 shows such a case of $\Delta \tau=10$ msec,
in which neurons from $j=1$ to 10 fire as the
pattern {\boldmath $\xi^{(1)}$} but not in a coherent way.
The time dependent overlap defined by Eq.(17) oscillates
with $0.88 \leq m(t) \leq 1.0$, as is shown in the upper
part of Fig. 10.  

The critical width of the time-delay distribution,
$\Delta \tau_c$, above which 
the retrieval of the stored patterns is failed, depends on the average
time delay of $\tau_d$. 
Figure 11 shows $\Delta \tau_c$ for three cases of 
$(f, \:\alpha)$=(0.06, 0.30), (0.06, 0.10) and (0.10, 0.10).
We should note that the time delay has the lower bound of 
7 msec below which the retrieval is not possible
because of the absolute refractory period of a HH neuron [Fig.1],
while it has no upper bounds.
For both $\alpha$=0.30 and 0.10,
the critical width for $f=0.06$
has a maximum value of $\Delta \tau_c=13$ msec
at $\tau_d=15$ msec, and $\Delta \tau_c \sim 8$ msec at 
$\tau_d \geq 25$ msec.
The critical width for $f=0.10$ is 
$\Delta \tau_c=13$ msec at $\tau_d \simg 15$ msec,
which is larger than that for $f=0.06$.
Our HH neuron network is fairly robust against the distribution
of the time delays in couplings.

\vspace{0.5cm}
\begin{center}
{\bf IV. DISCUSSION AND CONCLUSION}
\end{center}

Our simulations have demonstrated that neural networks consisting
of spiking HH neurons may show a fairly good performance as
an associative memory with the memory capacity of
$\alpha_c \sim 0.4-2.4$ for $f \sim 0.04-0.10$ [Fig. 5].
It is worth to compare the storage capacity of
our HH neuron networks with that of the {\it rate-code} model
in which the dynamics of the firing rate $x_j(t)$ is given by
\begin{equation}
x_{j}(t+1) = \Theta(\sum_{k(\neq j)} (W_{jk}-\nu) \: x_k(t)-\theta).
\end{equation}
In Eq.(20) $W_{jk}$ is determined by the Willshaw-type rule
given by Eq.(2), and
$\nu \:(0 < \nu < 1)$ and $\theta$ are the inhibitory coupling 
and the threshold, respectively, which are introduced 
to suppress cross-talk noises \cite{Golomb90}-\cite{Brunel94}. 
The storage capacity $\alpha_c$ depends on $f$, $\nu$ and $\theta$.
The calculated $\alpha_c$ 
of this {\it rate-code} model is shown as a function of $f$ 
by the solid curve in Fig. 12, where 
parameters of $N=100$, $\nu=0.8$ and $\theta=0.5$ are adopted \cite{Note3}.
We get the memory capacity of $\alpha_c \sim 0.4-3.7$ for
$f \sim 0.04-0.10$, which is comparable to that of
our HH neuron network.

For a comparison,
we have repeated calculations of the storage capacity
of the {\it rate-code} model mentioned above, employing the 
conventional Hebb-type rule given by Eq.(2)
instead of the Willshaw-type one given by Eq.(3). 
The calculated $f$ dependence of the
memory capacity is plotted by
the dashed curve in Fig. 12.
The storage capacity becomes $\alpha_c \sim 0.2-1.5$ for
$f \sim 0.04-0.10$, which is about a half of the values
in the {\it rate-code} model with the Willshaw-type synapses.
The critical storage capacities in the Hopfield model
with Willshaw-type [Eq.(2)] and Hebb-type synapses [Eq.(3)]
increase as the activity $f$ decreases, which is in agreement
with the theoretical analysis \cite{Tsodyks88}-\cite{Vicente89}.

We now examine the following two assumptions
adopted in Eq.(10) for the interaction term
in our HH neuron network:

(i) the Willshaw-type rule for weights of excitatory synapses and
the uniform inhibitory synapses [Eq.(11)], and

(ii) the uni-directional function, $G(x)$ [Eq.(12)]
for dendrite processing.

\noindent
We have employed the assumption (i) to enhance the storage capacity.
In order to show this explicitly, 
we calculate the $f$ dependence of the
storage capacity $\alpha_c$ when the Willshaw-type learning rule
given by Eq.(11) in our HH neuron network 
is replaced by the conventional Hebb-type one:
\begin{equation}
g_{jk} = g_{exc} \:
\sum_{\mu=1}^{P} \: \xi_{j}^{(\mu)} \: \xi_{k}^{(\mu)}.
\end{equation}
The calculated storage capacity $\alpha_c$ is plotted 
by the solid curve in Fig. 13 as 
a function of the average activity $f$.
The memory capacity is $\alpha_c \sim 0.03-0.14$ for
$f \sim 0.04-0.10$, which is much smaller than the corresponding value
of $\alpha_c \sim 0.4-2.4$ obtained with Willshaw-type rule [Fig. 5].
This clearly shows that the storage capacity
with the Willshaw-type rule [Eq.(11)] is enhanced by a factor 
of ten.
The reason why the solid curve in Fig. 13 has a discontinuous change at $f \sim 10$,
is not clear at the moment. The solid curve loosely follows the dotted
curve expressing $1/(M \; {\rm log} M)$, which is proportional 
to $(-1/f \; {\rm log} f)$ at $f \rightarrow 0$.

The assumption (ii) on $G(x)$ has been introduced to prevent firings for
inhibitory (negative) inputs.
This function $G(x)$ may be regarded to mimic the {\it inward rectification}
of dendrites \cite{Note4}, which make the response of a neuron to
hyperpolarizing outward currents much worse than that
to depolarizing inward currents.
If the uni-directional
function: $G(x)=x \: \Theta(x)$ given by Eq.(12) is replaced by
the linear one: $G(x)=x$, HH neurons fire for both excitatory and
inhibitory inputs,
and then the neural networks cannot
retrieve the stored patterns.
In this case, the weight of inhibitory synapses, 
$g_{jk}^{inh}$, in Eq.(10)
is properly determined by a learning rule
depending on the stored patterns, as 
Lytton \cite{Lytton98} has proposed for the feedforward HH network. 
Such a procedure, however,
yields a poor performance as an associative memory
working only in a narrow parameter range \cite{Lytton98}.
It would be necessary that the assumptions (i) and (ii) adopted in our
HH neuron network are examined and sought physiologically.

It has been reported that a large variability ($c_v=0.5 \sim 1.0$) is
observed in spike trains of non-bursting cortical neurons
in visual V1 and MT of monkey \cite{Softky92}, which is in strong contrast
with a small $c_v \;(= 0.05 \sim 0.1)$ in motor neurons \cite{Calvin68}. 
There have been much discussions how to understand 
the observed large variability \cite{Shadlen94}-\cite{Brown99}:
a balance between excitatory and
inhibitory inputs \cite{Shadlen94},
the high physiological gain in the plot of 
input current vs. output frequency \cite{Troyer98}, 
correlation fluctuations in recurrent 
networks \cite{Usher94},
the active dendrite conductance \cite{Softky95},
input ISIs with the distribution of a slow-decreasing tail \cite{Feng98},
and input ISIs with large $c_v$ \cite{Hasegawa00}\cite{Brown99}.
Based on our simulations,
we discuss the behavior of the interspike interval (ISI) 
of output spike trains which is  defined by 
\begin{equation}
T_{ojm}=t_{ojm+1}-t_{ojm},
\end{equation}
where $t_{ojm}\; (m=1,2,..)$ is the firing time of the neuron $j$ [Eq.(9)].
Histograms shown in Figs. 14(a)-14(c) express the distribution of ISIs
of {\it all} HH neurons for $0 \leq t \leq 500$ msec.
Figure 14(a) denotes the distribution of output ISIs of a successful retrieval 
with the initial overlap of $m_i=0.84$, which have been
discussed in Sec. 3.2: the time course of its firings of neurons is plotted
in Fig. 6(a) with parameters of
$f=0.20$, $\alpha=0.10$ and $\tau_d=10$ msec.
The mean and root-mean-square (RMS) values of the output ISI are
$\mu_o=11.98$ and $\sigma_o=1.42$ msec, which yield the
dimensionless {\it variability} of $c_v \equiv \sigma_o/\mu_o=0.12$.
The histogram of Fig. 14(b), on the other hand, shows the
distribution of ISI for a failed retrieval case with a smaller
initial overlap of $m_i=0.68$: the time course of neuron 
firings are plotted in Fig. 6(b).
We note that ISIs distribute up to 30 msec, 
obtaining $\mu_o=13.98$, $\sigma_o=5.69$ msec, and $c_v=0.41$.
The histogram of Fig. 14(c) expresses also a failed case
with more smaller $m_i=0.56$, which leads to 
$\mu_o=16.17$, $\sigma_o=17.76$ msec and $c_v=1.10$.
Note that there is a small distribution of ISI even 
at $30 < T_o < 60$ msec, where plotted result is 
magnified by a factor of ten.
Repeating these calculations, we obtain, in Fig. 15,
the $m_i$ dependence of $\mu_o$, $\sigma_o$ and $c_v$  
(for $f=0.20$, $\alpha=0.10$ and $\tau_d=10$ msec).
For $m_i > 0.9$, we get $\mu_o=11.76$ and $\sigma_o \sim 0$ msec, 
then $c_v \sim 0$.
When $m_i$ is more reduced, $\mu_o$ is increased because
ISIs with $T_o > 20$ msec appear, which make $\sigma_o$ and $c_v$ increase. 
Figure 15 shows that $c_v$ is larger for smaller $m_i$, and
that the variability 
becomes considerable ($c_v \sim 0.5$) even for successful retrievals 
with $m_i > 0.7$ [Fig. 6].
The situation is similar when a retrieval is 
worsen because of spread distributions
of time delays in couplings.  
For example, in the case having been shown in Fig. 10, where we plot
the time course of firings for $\Delta \tau=10$ msec 
with the parameters of
$f=0.10$, $\alpha=0.10$, $\tau_d=10$ msec and $m_i=1.0$,
a considerable value of
$c_v=0.45$ is obtained which arises from $\mu_o=18.61$ and 
$\sigma_o=8.31$ msec.
Our simulations show that the variability of ISIs
may be considerable while neurons are trying to retrieve the stored memory. 
We hope that results of our simulations might have some relevance
to the observed large variability.

To summarized, taking account of the two issues on the Hopfield-type
neural networks raised in the Introduction,
we have proposed the 
biologically plausible network consisting of spiking
HH neurons with the Willshaw-type synapses, whose simulations 
demonstrate its storage capacity to be comparable to that of
the rate-code network. 
We may extend and modify our model into various directions.
For example, it is straightforward to apply our method to the
feedforward hetero-associative memory of temporal-code
Anderson-Kohonen-type networks.
When the excitatory synapse given by Eq.(11)
is changed as
\begin{equation}
g_{jk}^{exc} = g_{exc} \; \Theta
(\sum_{\mu=1}^{P-1} \: \xi_{j}^{(\mu+1)} \: \xi_{k}^{(\mu)}),
\end{equation}
the patterns of $\mu=1-P$ can be retrieved in a temporal sequence. 
This shows a possibility of the storing and 
retrieval of temporal patterns.
The synapse given by Eq.(23) corresponds to the asymmetric coupling
in the Hopfieled model \cite{Sompolinsky86}.
It has been shown that the Hopfieled model with asymmetric couplings
shows intrigue dynamics like chaos. 
In the present paper, we have taken no account of noises.
Quite recently, 
Kanamaru and Okabe \cite{Kanamaru00} have investigated the effect of noises
on an associative memory of the FN neuron network,
showing that a retrieval of the stored patterns may be
improved by noises, just like the stochastic resonance. 
It would be the case also in our HH neuron network.
We cannot, however, draw any definite conclusion until
performing simulations because their leaning rule
given by Eq.(1) with $a=0$ and $b=f$
is different from ours given by Eq.(11).
It is well known that temporal-code neurons are more efficient
in data processing than the rate-code neurons
because the former can carry much information than 
the latter \cite{Rieke96}.
It is expected to be true also in the memory function.
We suppose that there could be alternative, more efficient 
mechanisms, by which real temporal-code networks might be operating
for learning and memory.

\section*{Acknowledgements}
This work is partly supported by
a Grant-in-Aid for Scientific Research from the Japanese 
Ministry of Education, Science and Culture.



\begin{figure}
\caption{
Responses of the membrane potentials, $V$ (solid curve), 
of a HH neuron to (a) excitatory and (b) inhibitory external 
currents, $I^{ext}$ (dashed curve):
a HH neuron fires not only by excitatory input but also by
inhibitory inputs.
}
\label{fig1}
\end{figure}

\begin{figure}
\caption{
The time course of (a) input $U_{i1}$,
(b) output $U_{o1}$, 
(c) the post-synaptic current $I_1\;(=I_1^{ext}+I_1^{int})$ and
(d) the membrane potential $V_1$ of the neuron $j=1$ 
when an input pattern is injected at $t=0$
with $\alpha=0.3$, 
$f=0.10$, $g_{exc}=0.30 \; {\rm m S/cm}^2$ and $\lambda=0.8$;
the vertical scale is for $V_1$ only, and $I_1^{int}$, $U_{o1}$ and $U_{i1}$
are shown in arbitrary units.
}
\label{fig2}
\end{figure}

\begin{figure}
\caption{
The time course of firings of neurons 
for (a) $\alpha=0.30$ and (b) $\alpha=0.50$ 
with $f=0.10$, $g_{exc}=0.30 \; {\rm m S/cm}^2$ and $\lambda=0.8$;
dots express firings of neurons numbered by index shown in the 
vertical scale:
the upper part shows
the relevant overlap of $m(t)$.
}
\label{fig3}
\end{figure}

\begin{figure}
\caption{
The storage capacity $\alpha_c$ as a function of 
$\lambda\:(=g_{inh}/g_{exc})$ for $f=0.06$ and 0.10
with $g_{exc}=0.3 \; {\rm m S/cm}^2$ (solid curve) and
$g_{exc}=0.5 \; {\rm m S/cm}^2$ (dashed curve).
}
\label{fig4}
\end{figure}

\begin{figure}
\caption{
The storage capacity $\alpha_c$ as a function 
of the average activity $f$.
The solid curve denotes the calculated results
and the dotted curve expresses $10/(M \: {\rm log} M)$,
error bars denoting the standard deviations
for ten samples.
}
\label{fig5}
\end{figure}

\begin{figure}
\caption{
The time course of neuron firings for an injection of 
imperfect input patterns with initial overlaps of (a) $m_i=0.84$
and (b) $m_i=0.68$ for $f=0.20$ and $\alpha=0.10$.
}
\label{fig6}
\end{figure}

\begin{figure}
\caption{
The time dependence of the overlap $m(t)$
for various $m_i$ with $f=0.20$ and $\alpha=0.1$.
}
\label{fig7}
\end{figure}

\begin{figure}
\caption{
The storage capacity $\alpha_c$ as a function 
of the initial overlap $m_i$ 
for $f=0.10$ (solid curve) and $f=0.20$ (dashed curve).
}
\label{fig8}
\end{figure}

\begin{figure}
\caption{
The storage capacity $\alpha_c$ as a function 
of a width of the time-delay distribution, $\Delta \tau$,
for $\tau_d=10$ msec.
The solid and dashed denotes the calculated results
for $f=0.06$ (solid curve) and $f=0.10$ (dashed curve), 
respectively.
}
\label{fig9}
\end{figure}

\begin{figure}
\caption{
The time course of firings of neurons for 
$f=0.10$, $\alpha=0.10$,
$\Delta \tau=10$ and $\tau_d=10$ msec: 
the upper part shows
the relevant overlap of $m(t)$ (see text).
}
\label{fig10}
\end{figure}

\begin{figure}
\caption{
The critical width, $\Delta\tau_c$, of the time-delay distribution
as a function of $\tau_d$ 
for $(f, \; \alpha)=(0.06, 0.30)$ (solid curve),
(0.060, 0.10) (dotted curve) and (0.10, 0.10) (dashed curve). 
Calculations are made
with a step of $\Delta \tau=1.0$ msec for a given $\tau_d$.
}
\label{fig11}
\end{figure}

\begin{figure}
\caption{
The storage capacity $\alpha_c$ as a function 
of the average activity $f$ 
of the {\it rate-code} neuron network with
the Willshaw-type synapses (solid curve) and 
the Hebb-type synapses (dashed curve),
the dotted curve denoting $10/(M \:{\rm log} M)$ (see text).
}
\label{fig12}
\end{figure}

\begin{figure}
\caption{
The storage capacity $\alpha_c$ as a function 
of the average activity $f$ of the HH neuron network
with the Hebb-type synapses [Eq.(21)].
The solid curve denotes the calculated result
and the dotted curve expresses $1/(M \: {\rm log} M)$.
}
\label{fig13}
\end{figure}

\begin{figure}
\caption{
Histograms of ISIs of spike-train outputs in a retrieval process
with the initial overlap of (a) $m_i$ =0.84, (b) 0.68 and (c) 0.56
for $f=0.20$, $\alpha=0.10$ and $\tau_d=10$ msec,
the histogram at $T_o > 30$ msec of (c) being enlarged
by a factor of ten.
The time course of neurons firings for (a) and (b)
are plotted in Fig. 6(a) and 6(b), respectively.
}
\label{fig14}
\end{figure}

\begin{figure}
\caption{
The average ($\mu_o$, thin solid curve) and 
RMS values of ISIs ($\sigma_o$,  dashed curve),
and the variability ($c_v$, bold solid curve)
as a function of the initial overlap ($m_i$)
for $f=0.20$, $\alpha=0.10$ and $\tau_d=10$ msec,
the left (right) vertical scale being for $\mu_o$ and
$\sigma_o$ ($c_v$).
Error bars denote the standard deviation of $c_v$,
and those for $\mu_o$ and $\sigma_o$ are not shown 
for illegibility of the figure.
}
\label{fig15}
\end{figure}

\end{document}